\documentclass[preprint,aps,showpacs]{revtex4}

\usepackage{graphicx}
\usepackage{dcolumn}
\usepackage{bm}
\begin{document}

\title{
First-principles study of the atomic and electronic structure of the Si(111)-(5$\times$2)-Au surface reconstruction}

\author{Sampsa Riikonen}
\email{swbriris@sc.ehu.es}
\affiliation{Departamento de F\'{\i}sica de Materiales,
Facultad de Qu\'{\i}mica, Universidad del Pa\'{\i}s Vasco (UPV/EHU),
Apdo. 1072,
20080 San Sebasti\'an, Spain}
\affiliation{Donostia International Physics Center (DIPC),
Paseo Manuel de Lardizabal 4, 20018 San Sebasti\'an, Spain}

\author{Daniel S\'anchez-Portal}
\email{sqbsapod@sc.ehu.es}
\affiliation{
Unidad de F\'{\i}sica de Materiales,
Centro Mixto CSIC-UPV/EHU, Apdo. 1072,
20080 San Sebasti\'an, Spain}
\affiliation{Donostia International Physics Center (DIPC),
Paseo Manuel de Lardizabal 4, 20018 San Sebasti\'an, Spain}

\date{\today}

\begin{abstract}
We present a systematic study of the atomic and
electronic structure of the
Si(111)-(5$\times$2)-Au reconstruction
using first-principles 
electronic structure calculations based on the
density functional theory. 
We analyze the structural models proposed 
by Marks and Plass [Phys. Rev. Lett. {\bf 75}, 2172 (1995)],
those proposed recently by Erwin 
[Phys. Rev. Lett. {\bf91}, 206101 (2003)],
and a completely 
new structure that was found during our structural optimizations.
We study in detail the energetics and the structural and electronic
properties of the different models. For the two most stable models,
we also calculate the change
in the surface energy as a function
of the content of silicon adatoms
for a realistic range of concentrations.
Our new model is the energetically most favorable in the range
of low adatom concentrations, 
while Erwin's  ``5$\times$2" model
becomes favorable 
for larger adatom concentrations. The crossing between
the surface energies of 
both structures is found 
close to 1/2 adatoms per 5$\times$2 unit cell, i.e. near 
the maximum adatom coverage
observed in the experiments.
Both models, the new structure and Erwin's ``5$\times$2" model,
seem to provide a good description of many of the available experimental
data, particularly of the angle-resolved photoemission measurements.

\end{abstract}


\maketitle

\section{Introduction}

The low-energy electronic spectrum of a one-dimensional metal
is dominated 
by collective spin and charge 
excitations~\cite{Giamarchi04,luttinger1,luttinger2}. 
This is in contrast with the behavior of typical metals, that 
can be understood in terms of independent particle-like excitations 
usually called quasiparticles.
These predictions are clear and well
established. However, the observation 
of the Luttinger liquid behavior in real systems 
has proven quite elusive. 
One of the reasons for this might be that one-dimensional
metals are in principle 
unstable with respect to the Peierls distortion that 
drives them into an insulating ground state~\cite{peierls}.
A possible route to avoid this limitation is the fabrication
of metallic chains absorbed on surfaces. The hope is 
that the rigidity of the substrate will make the 
energy cost for the structural distortions too large
and, therefore, the one-dimensional chains
would remain metallic. Semiconductor surfaces
are specially attractive for this purpose. 
The existence of an energy gap prevents the coupling
of the electronic  
states of the chain in the vicinity 
of the Fermi level with the substrate.
The one-dimensional character of these states 
is thus preserved.
It is in this context, that the fabrication 
of monatomic wires of metal
atoms on silicon substrates has attracted much attention in
recent years. 

Monatomic wires of gold atoms are spontaneously formed 
on flat and vicinal Si(111) surfaces after the deposition
of gold in the sub-monolayer regime 
(see Ref.~\onlinecite{himpsel_review}
and references therein). Of particular
interest are the vicinal substrates, where
gold wires run parallel to each step-edge and
the coupling between the chains, and 
thus the one-dimensional
character of the electronic states, 
can in principle 
be controlled by changing the miscut angle.
Examples of these systems that have attracted much attention
in the last few years are the 
Si(557)-Au~\cite{Segovia,Losio01,Ahn03,daniel1,robinson,daniel2,
daniel3} and the 
Si(553)-Au~\cite{Crain03,himpsel_review,Crain05,sampsa05}
reconstructions. 
The photoemission spectra of these surfaces close to
the Fermi energy are dominated 
by bands with a strong one-dimensional character
that exhibit several interesting phenomena including 
peculiar splittings associated with 
the spin-orbit interaction~\cite{daniel3} 
(although first interpreted as signature
of spin-charge separation~\cite{Segovia}), 
fractional fillings~\cite{Crain03} and metal-insulator
transitions~\cite{Ahn03}.
The analogous to these one-dimensional
structures in the case of the 
flat Si(111) surface is the so-called 5$\times$2 reconstruction.

The deposition of gold in the monolayer (ML) range on the flat
Si(111) surface results in a variety of phases \cite{5x2_phase},
such as $\sqrt{3}\times\sqrt{3}$R30$^{o}$, 1$\times$1 and 5$\times$2.
The 5$\times$2 phase occurs at $\sim$0.4~ML gold coverage~\cite{bauer}.
It was first discovered about 
thirty years ago~\cite{5x2_leed1,5x2_leed2,5x2_leed3} 
and has been investigated using many experimental techniques since then.
This includes 
low energy electron diffraction (LEED) 
studies~\cite{5x2_leed1,5x2_leed2,5x2_leed3},
x-ray diffraction~\cite{x_ray_5x2_2} and
x-ray standing wave analysis~\cite{x_ray_5x2}, 
scanning tunneling microscopy (STM)~\cite{stm_5x2,stm_5x2_2,hhh},
angle resolved photoemission spectroscopy 
(ARPES)~\cite{arpes_1,5x2_bands,1d_states,arpes_latest,himpselnewARPES04}
and inverse photoemission~\cite{arpes_2}, and
high resolution electron microscopy (HREM) combined with 
heavy-atom holography~\cite{marks_plass}.

Already the first structural
models, based on LEED measurements,
considered two atomic gold chains per 5$\times$2
unit cell 
running in parallel~\cite{5x2_leed2,5x2_leed3}.
This was later confirmed by HREM~\cite{marks_plass} and seems to be
firmly established. The gold chains run along the $[\bar{1}10]$
and equivalent directions (parallel to the $\times$2 periodicity 
of the unit cell).
Therefore, three different domains are possible for the 5$\times$2
reconstruction on the flat Si(111)
surface. Single-domain surfaces, necessary for ARPES,
can be fabricated using 
vicinal surfaces with a slight cut-off  
angle~\cite{himpsel_review,stm_5x2_2}.
The presence of one-dimensional structures in 
this reconstruction has also been confirmed
by the ARPES studies. Early studies found a 
strong anisotropic signal near the
Fermi level \cite{arpes_1,arpes_2}, but no evidence of Fermi-level
crossing for this band was found \cite{arpes_2}.
Later studies at low temperature found a
one-dimensional band with a strong dispersion along
the direction of the gold chains~\cite{5x2_bands,1d_states}. 
The top of this band
appears near the 5$\times$2 zone boundary and disperses
downward, reaching its minimum close the 5$\times$1 
zone boundary.
This band has been reported to change its dimensionality from strongly
one-dimensional near the Fermi energy to two-dimensional 
at lower energies~\cite{1d_states}. 
In these studies a gap of $\sim$0.3~eV was also identified for
this band. The presence of this gap 
and its apparent closing with increasing 
temperature was 
related to a Peierls instability~\cite{5x2_bands,1d_states}.
More recent ARPES 
results~\cite{arpes_latest,himpselnewARPES04}, both at low and room 
temperatures, have been able to identify some additional
surface bands. However, the metallic or semiconducting
character of the surface is still a matter of debate. In 
fact, it
has been proposed that the metallic or semiconducting 
character can depend on 
the concentration of silicon 
adatoms~\cite{erwin03,himpselnewARPES04}, and even 
that semiconducting and metallic segments can
alternate along the gold chains in the surface~\cite{Yoon04}.

The STM images are characterized by the presence of 
bright, irregular protrusions~\cite{stm_5x2,hhh},
and ``Y"-shaped features~\cite{stm_5x2_2,Yshaped} with 
a well defined
orientation respect to the underlying substrate.
The protrusions have been established to be silicon 
adatoms~\cite{adatoms}, which are present on the surface
with an optimum
coverage close to 1/4 adatoms per 5$\times$2 unit cell.
                                                                                
In spite of all these experimental studies, the structure 
of the Si(111)-(5$\times$2)-Au
reconstruction has not been completely established yet.
Earlier structural models only 
considered the adsorption sites of the 
gold atoms. Many of them could be ruled out on the bases of more 
detailed STM studies~\cite{stm_5x2} and the knowledge of the exact 
gold coverage~\cite{bauer}. A few more refined
models exist~\cite{marks_plass,hhh}. They consider both 
the position of the gold atoms on the 
substrate and the rebonding of the silicon atoms in 
the surface layer.
Probably the most 
detailed structural model proposed to date is the 
one by Marks and Plass (MP)~\cite{marks_plass}.
The MP model is based on a combination 
off-zone HREM, transmission electron diffraction
and heavy-atom holography data.

The first theoretical studies
using first-principles electronic structure calculations 
appeared only quite recently.
This is due to the complicated structure
and the large unit cell of the Si(111)-(5$\times$2)-Au
reconstruction.
Kang and Lee~\cite{5x2_comp} studied the 
MP and the Hasegawa-Hosaka-Hosoki (HHH)~\cite{hhh} models 
using 
density functional theory. Their main conclusion is that
both models fail to 
reproduce some of the key features
of the STM images and the 
experimental band structures. Using a similar methodology,
Erwin~\cite{erwin03} proposed and studied new structures
which are characterized by the presence of the 
so-called honeycomb-chain silicon structure~\cite{honeycomb}.
One of these models (the so-called ``5$\times$2" model)
seems to fulfill many of the constraints imposed
by the empirical evidence. An interesting point raised 
by Erwin is that of the crucial role played by 
silicon adatoms in the 
stabilization of the different structures. 
According to Ref.~\onlinecite{erwin03},
the surface energy of Erwin's ``5$\times$2" model
is minimized for an optimum adatom
coverage in agreement with recent experimental reports~\cite{adatoms}.
For lower adatom coverages other structures compete in 
stability.
This is a very interesting result which, however, is 
based on approximate calculations.
Due to the large size of the supercells necessary to 
simulate explicitly the effect of the different adatom concentrations,
Erwin assumed that the main role played by the adatom is to dope the
surface with electrons. He then analyzed the behavior of the total
energy as a function of the number of extra-electrons in the substrate,
obtaining a minimum for $\sim$0.25 electrons per 5$\times$2 unit cell.

In this work, we present a comprehensive study of the atomic and
electronic structure of different models of the 
Si(111)-(5$\times$2)-Au reconstruction
using electronic structure calculations based on the 
density functional theory. We have used 
two different methodologies, the 
SIESTA code~\cite{siesta1,siesta2,siesta3}
using a basis set of localized atomic orbitals and 
the VASP code~\cite{vasp1,vasp2} using
a basis set of plane-waves. We analyze the MP model~\cite{marks_plass},
the models proposed by Erwin~\cite{erwin03},
and a new model that we found during our structural optimizations.
We study in detail the energetics and the structural and electronic
properties of the different models. We also calculate 
the change in 
the surface energy as a function
of the content of silicon adatoms
for the two most stable models. In order to do so, we
perform calculations
for large supercells containing 
realistic concentrations of adatoms: 5$\times$4, 5$\times$6,
and 5$\times$8 supercells.
Our new model is the most favorable in the range
of low adatom concentrations, while Erwin's ``5$\times$2" model
becomes favorable
for larger adatom concentrations. The crossing between the
surface energy of  
both structures occurs
close to 1/2 adatoms per 5$\times$2 unit cell, i.e. near 
the maximum
adatom concentration observed in the experiments.
Both models, our new structure and Erwin's ``5$\times$2" model,
seem to provide a good description of most of the experimental
data. Particularly, we find a general  
agreement between the calculated and measured 
band structures along
the direction parallel to the gold chains.

\section{Calculation method}

Most of our calculations have been performed using the 
SIESTA~\cite{siesta1,siesta2} code. 
We have used here the local approximation (LDA) 
to the density-functional theory\cite{KS,CA,PZ}
(the generalized gradient approximation (GGA) has been also used
for a few test calculations),
Troullier-Martins pseudopotentials\cite{TM}, and a
basis set of numerical atomic orbitals obtained from the solution
of the atomic pseudopotential at slightly excited 
energies~\cite{SankeyNiklewski,siesta1,junquera,siesta2}.
We have used an {\it energy shift}~\cite{junquera,siesta2} of
200~meV. The corresponding radii are 5.3, 6.4 and 6.4 a.u.
for the $s$, $p$ and $d$ orbitals of Si, 6.2, 6.2 and 4.5 in the 
case Au, and 5.1 for the $s$ and $p$ states of H.
The bidimensional Brillouin-zone (BZ) sampling~\cite{MonkhorstPack} 
contained
4$\times$4 points for the
5$\times$2 unit cell (and a consistent sampling for other
cells~\footnote{More precisely, for the 5$\times$4 supercells
we have checked that k-samplings with 4$\times$2 and 4$\times$4 points
produce
almost identical results. k-samplings with
4$\times$3 and 4$\times$2 points have been used, respectively, for 
the 5$\times$6 and 5$\times$8 supercells.}) 
and the fineness of the real-space grid used
to compute the Hartree and exchange-correlation contributions
to the total energy and Hamiltonian matrix elements was
equivalent to a 100~Ry plane-wave cutoff. 
This guarantees the convergence of the total
energy, for a given basis set, within $\sim$20~meV/Au
($\sim$0.5~meV/\AA$^2$).

We modeled the surface using a finite slab, 
similar to that depicted in
Fig.~\ref{fig:slab}. For most calculations 
the slabs contained three
silicon bilayers (the one at the surface and two underlying
silicon bilayers) plus an additional layer of hydrogen atoms 
to saturate the silicon atoms in the bottom of the slab. 
This removes the surface bands associated to the bottom surface
from the 
energy-range of interest, i.e. from the band-gap region. 
We have checked the convergence of the results using
thicker slabs for the most stable structural models of the
surface.
We use periodic boundary conditions in all three directions.
A vacuum region of 15~\AA\ 
ensures negligible interactions between neighboring slabs.
During the structural relaxations
the positions of the silicon atoms in the bottom layer 
were kept at the bulk ideal positions. Unless otherwise stated all
other degrees of freedom were optimized until all the components
of the residual forces were smaller than 0.04~eV/\AA.
To avoid artificial stresses the lateral 
lattice parameter was adjusted
to the theoretical bulk value calculated using similar approximations
to those utilized in the slab calculations, 
i.e. the same basis set and grid cutoff, and a consistent
k-sampling.
The values are, respectively, 
5.48 and 5.42~\AA\ with double-$\zeta$ (DZ)
and double-$\zeta$ polarized (DZP) basis, to be compared with the 
experimental value of 5.43~\AA.

Due to the large number 
of atoms ($\sim$70 atoms for typical 
slabs and up to 273 for the largest ones)
and to the need to perform
geometrical optimizations for
many different structural models, we have decided to use a DZ
basis set for silicon in
most of our calculations. This basis set includes two different functions (i.e.
two different radial shapes) to represent 
the 3$s$ orbitals of Si, and another
two for the 3$p$ shell. We have tested the performance 
of this basis set and the other
parameters of our calculations using
the well known 
Si(111)2$\times$2 adatom 
reconstruction as a test. Our results using a DZ basis
and a slab containing 
three bilayers reproduced very well the results
for the geometry
and energetics given by previous calculations\cite{Vanderbilt}. 
In fact, using a more complete DZP
basis set, which 
includes a shell of $d$ orbitals, or increasing 
the thickness of the slab (up to 
five bilayers) did not change appreciably the results.

A scalar-relativistic pseudopotential similar to that utilized 
in Ref.~\onlinecite{au1} have been used for gold.
The gold basis set included
double and polarized
6$s$ orbitals (i.e. two different radial shapes to 
describe 6$s$ orbital plus a
6$p$ shell) and a single 5$d$ shell. We refer to this basis 
set as DZPs-SZd. This basis set was already 
used in calculations for several gold clusters, where it was
shown to lead to results in very good agreement with 
more complete basis sets~\cite{PRBAuclusters}.
We have also tested that these basis set and pseudopotential 
yield to the correct bulk lattice parameter and band structure. 

In this work we study the relaxed structures 
and the energetics of 
several models of
the Si(111)-(5$\times$2)-Au
surface reconstruction.
The energy differences between different models
are of key importance since we would like to determine the most
plausible structures.
Whenever it is necessary to compare the energies of 
structures containing
different numbers of silicon atoms, the silicon
chemical potential is set to the total energy of bulk
silicon at the equilibrium lattice parameter. This choice is justified
by the fact that the surface should be in equilibrium with the bulk.
A summary of our results can be found in Table~\ref{tab:systems}.
One can see that the relative energies are quite small
in some cases. However,
they are larger than the estimated error bar for the total energy 
(see above). Furthermore, 
the relative energies usually 
exhibit a faster convergence  
than the total energy of a single structure.
It is also necessary to check the convergence of the results
as a function of the slab thickness
and the completeness of the basis set.
Table~\ref{tab:systems2} shows the results of these tests
for the most stable structures. In one case, the 
slab thickness was increased 
by one silicon bilayer while, in the other,
a DZP basis set was used for the silicon 
atoms.
In both cases the systems were relaxed.
The results are quite stable against
the change of the slab thickness. In particular, the
energy order of the structures 
is not changed and the 
variation of the relative surface energies is 
smaller than $\sim$0.5~meV/\AA$^2$ in all the cases. 
The variations with the size of the basis set are somewhat larger.
From the results in Table~\ref{tab:systems2}
we can estimate an error bar smaller than
2~meV/\AA$^2$ for the relative surface energies
of the different structures calculated using SIESTA.

In order to check the accuracy of
our predictions we decided to perform 
calculations for some of the systems
with another electronic structure code that utilizes a 
different methodology. We used the 
VASP code~\cite{vasp1,vasp2} for this purpose. 
We used projected-augmented-wave
potentials and a  well converged
plane-wave basis set with a cutoff of 312~eV. All
structures were relaxed (the equilibrium lattice parameter
of bulk silicon obtained with VASP is 5.41~\AA).
In Table~\ref{tab:systems2} we can see some of the results
obtained with VASP. They are in good agreement with 
the SIESTA results, especially with those obtained with 
the more complete 
DZP basis set. The order between our more stable 
models is preserved, although the energy difference
is somewhat decreased. In particular, the new
structural model found in the present 
work (model N in Tables~\ref{tab:systems} and \ref{tab:systems2}) 
is confirmed to be the most stable surface reconstruction
between those studied here. It is also interesting to note that
the energy associated
with the addition and removal of adatoms 
for a particular structural
model seems to be quite
independent of the details of the calculation.  

The surface BZs of the studied systems are shown in 
Fig.~\ref{fig:bzs}~(a).
For the 5$\times$1 system the BZ is a stretched hexagon while,
for the remaining periodicities, the hexagons are distorted. 
We plot the electronic band structures of the different models
along the $\Gamma$-
ZB$_{\times2}$-ZB$_{\times1}$-ZB$_{\times2}^{\prime}$
-M-$\Gamma$ line 
(see the dotted line in Fig.\ref{fig:bzs}~(b)).
The $\Gamma$-M path runs parallel to the gold wires in the 
surface, crossing the 5$\times$2 BZ through three different
regions. The M-$\Gamma$ line is perpendicular 
to gold wires. The surface/bulk and main atomic 
character of the different
bands is identified
by means of a Mulliken population analysis\cite{mulliken}.

Although a DZ basis is usually sufficient to obtain a quite
good description of the occupied electronic states and 
the relaxed geometries in silicon systems, the use of 
a more complete basis set is necessary to describe the 
unoccupied part of the band structure even at low energies. 
For this reason
all the band structures shown in the paper are calculated using
a DZP basis set and slabs containing three underlying silicon bilayers
(even if the relaxed geometry is obtained from a calculation
using a DZ basis and/or a thinner slab). 

Finally, the Scanning Tunneling Microscopy (STM) images are 
simulated using the theory of Tersoff and Hamann\cite{stm_theory}.

\section{Results}

In this section we present our results for the 
different models of the Si(111)-(5$\times$2)-Au surface.
We first focus on the energetics, 
relaxed geometries, and 
the electronic band structures. 
We then turn our attention to the effect of the different 
silicon adatom contents and the simulated STM images, which we
only analyze in detail for the most
stable structural models.
A summary of the relative energies of the calculated configurations,
accompanied with a brief description of each of them, 
can be found in Table~\ref{tab:systems}.

Before starting with the description of the results,
it is interesting to point out some brief comments about 
the concentration of silicon adatoms
on the Si(111)-(5$\times$2)-Au surface.
A detailed study of the equilibrium situation 
has recently been performed by 
Kirakosian {\it et al.}~\cite{adatoms,5x2_corr} 
using STM.
Their results indicate that, at
equilibrium, only one
adatom site is occupied out of every four possible sites, corresponding to a
5$\times$8 adatom periodicity (if all the adatom sites were occupied 
we would recover a perfect 5$\times$2 periodicity).
The analysis of
the adatom-adatom correlation functions obtained from the STM images 
reveals a strong suppression of those configurations with 
small adatom-adatom distances,
a clear
maximum corresponding 5$\times$4 periodicities,
and a long range oscillatory 
tail~\cite{5x2_corr}.
This was interpreted in terms of a short range repulsion between adatoms
plus a long range interaction term. In Ref.~\onlinecite{adatoms}, 
Kirakosian and
collaborators showed that the density of adatoms can be
increased by depositing additional amounts 
of silicon reaching an almost perfect
5$\times$4 arrangement of the silicon adatoms.  
Further deposition of
silicon does not create 
a stable 5$\times$2 adatom structure. Instead the extra
silicon atoms decorate the step-edge 
of the terraces on the surface. These
observations seem to have at least two implications:
($i$) the optimal adatom concentration must be certainly lower than one adatom
per 5$\times$2 cell and, ($ii$) the structure of the reconstruction
must be stable against relatively large changes 
of the content of adatoms~\footnote{This observation does not
contradict the recent proposal by
Erwin~\cite{erwin03} that
a {\it minimum} adatom content
may be necessary to stabilize the observed structure over other competing
reconstructions.}
since the density of silicon
adatoms 
can be increased by a factor of two without, at least
apparently,
dramatic structural
changes~\cite{adatoms}.

A systematic study of the energetics of the 
surface as a function of the adatom
concentration by means 
of first-principles electronic structure calculations is quite complicated.
This is for two main reasons. First, the energies involved
are rather small, which implies the need of very well converged
calculations. A more serious limitation, however, is the necessity to
use large supercells consistent with the low adatom densities.
For this reason we have concentrated most efforts
in the two limiting cases, 
involving respectively 0 and 1 adatoms per 5$\times$2 cell.
The intermediate concentrations usually require drastic approximations.
For example, Erwin~\cite{erwin03} assumed
that the main effect of the adatoms in the Si(111)-(5$\times$2)-Au
surface is to dope the gold chains with electrons and studied
the energetics of the system as a function of this doping.
Here we go a step beyond and present explicit calculations  
for adatom contents down to 1/4 adatoms per 5$\times$2 cell, 
consistent with a 5$\times$8 periodicity,
which indeed can be reached in experimental conditions~\cite{adatoms}.
Due to the large size of these systems we limit
this study to our two most stable models, and only use 
the smaller DZ basis set.

\subsection{Marks and Plass model}

We start our investigation of the structure of the Si(111)-(5$\times$2)-Au 
surface using the model proposed by Marks and Plass~\cite{marks_plass} 
from experimental data obtained with heavy-atom holography and high resolution
electron microscopy. We use the label MP$^+$ for this structure (see 
Table~\ref{tab:systems}). The 
$+$ superscript indicates that the structure contains silicon 
adatoms saturating some
of the silicon dangling-bonds in the structure, what we call "conventional"
silicon adatoms.
A schematic view of this structure, as proposed 
in Ref.~\onlinecite{marks_plass}, can be found 
in Fig.~\ref{fig:slab}. It has to be taken into account that, due to 
the limitations of the experimental techniques, there are several uncertainties
and assumptions in this structure. Only the atomic coordinates within the
surface plane are accurate. The heights of the atoms over the substrate
are only approximately resolved. The experimental beam error in combination
with the size and complexity of the structure also limits the sensitivity 
to possible subsurface relaxations. As a consequence, the experimentally
proposed 
structure only considers the reconstruction of the outermost 
bilayer and contains
limited information about the registry between this surface bilayer and the 
underlying material.
It is necessary to eliminate these uncertainties before one can undertake any
serious study of the electronic structure of the MP$^+$ model.
In order to do this while preserving all the information originally present in 
the MP$^+$ proposal, we started our 
study by performing constrained relaxations  
of the structure.
The structure in Fig.~\ref{fig:slab} was relaxed using
following degrees of freedom: 
($i$) the height of the different layers and, 
($ii$) the lateral position of the surface layers 
with respect to the underlying bulk slab.
The grouping of the atoms in different layers 
given in Ref.~\onlinecite{marks_plass} only implies approximately
equal z-coordinates ( the z-axis is taken here along the surface normal).
For this reason, in a second step, the atoms were allowed to 
relax in the z-direction while keeping their coordinates within the xy-plane.
The resulting geometry 
preserves the bonding pattern of the original MP$^+$ proposal, and provides
a reasonable initial guess to start our search of the most stable models
by performing full structural optimizations.

We now consider in detail some of the structural patterns 
appearing in the MP$^+$
model of the surface.
For this analysis we find useful the comparison with the Si(557)-Au 
surface, a closely 
related reconstruction 
that has been quite well characterized during recent 
years~\cite{Segovia,Losio01,daniel1,robinson,daniel2,daniel3}.
The stepped Si(557)-Au is formed after the deposition of $\sim$0.2 monolayers
of gold on vicinal (111), with the misorientation chosen along the 
$[\bar{1}\bar{1}2]$ direction. The size and orientation of the 
terraces of the Si(557)-Au
represent an analogous to the flat 5$\times$2 unit cell but including a 
single silicon step~\cite{5x2_bands}. With half the gold
coverage than the Si(111)-(5$\times$2)-Au surface, the 
terraces of the Si(557)-Au contain
only one Au wire running parallel to the step edges. Gold
atoms occupy silicon substitutional positions in the surface layer.
This is supported both, by recent X-ray diffraction data~\cite{robinson},
and density functional calculations performed using a methodology
similar to the one utilized here~\cite{daniel2}, which provide a consistent
structural model of the surface. In particular, the highest stability of
the silicon substitutional sites for gold has been 
unambiguously demonstrated
by the {\it ab initio} calculations. For example, the substitutional 
site was determined to be at least 1~eV/Au more stable that 
adatom-like positions, where gold sits on the surface saturating one
of the silicon dangling-bonds, or even $\sim$0.5~eV/Au more favorable than 
the adsorption decorating the step edges~\cite{daniel2}. 
It seems, therefore, that the 
Au atoms on the Si(557)-Au surface exhibit a strong tendency towards 
three-fold silicon coordination. Gold atoms adapt to this situation
without much strain, with typical Si-Au distances 
only a few percents larger
than the bulk silicon bond length.  

In the light of these observations the bonding pattern of some of the
gold atoms in the MP$^+$ model (Fig.~\ref{fig:slab}) 
seems quite peculiar.
In particular, the gold
atoms in the chain situated at the left side of the "gold trench" 
(marked with an L in Fig.~\ref{fig:slab}) 
present a fourfold coordination.
They are connected to three silicon atoms 
within the surface layer and, additionally, 
to the silicon atom immediately below. Furthermore, the Si atoms
neighboring to the mentioned gold atoms (see atoms $a$ and
$a^\prime$ in 
Fig.~\ref{fig:slab})
present an unsaturated dangling bond which might be avoided with a
slight structural change.
                                                                                
It is interesting to note that the tendency of the gold atoms to occupy
silicon substitutional positions in the top most layer
cannot help to completely rationalize the
structure. A three-fold bonding pattern of the gold atoms is inherently
frustrated by the presence of a
surface dislocation. In the MP$^+$ model this dislocation is
located at the position of the right-hand gold wire (marked
with R in Fig.~\ref{fig:slab}).
Due to the change of the bonding sequence
there are not three unpaired silicon electrons available for
each of these Au atoms, but rather two.
Therefore, they do not occupy a normal three-fold position and are
quite likely to
be displaced from the initial symmetric positions after relaxation.
It should be noted that, in principle, the surface
dislocation can be moved to different locations.
In fact, we will see below that this provides a simple route
to generate alternative structural models of the surface.

The comparison between the structure of the Si(557)-Au
reconstruction\cite{robinson,daniel2} and the MP$^+$ 
model of Fig.\ref{fig:slab}
raises another interesting point. In the case of the Si(557)-Au
surface the silicon atoms in the proximity of the step edge
suffer a considerable rebonding. They form characteristic silicon
structure which has been 
identified with the so-called
"honeycomb chain" (HC) by several authors~\cite{daniel2,himpsel_review}. 
The presence of the silicon HC seems
instrumental to understand the stability of the Si(557)-Au and 
related reconstructions 
(see, for example, Ref.~\onlinecite{himpsel_review}).
The
silicon HC was initially proposed by Erwin and Weitering\cite{honeycomb} as
the main building-block of the (3$\times$1) reconstruction induced on Si(111)
by the
deposition of metals like Ag, Li, Na, K, Mg or Ba. The
HC structure represents a large disturbance from the usual bonding
pattern of silicon. The stability of the silicon HC stems fundamentally from
the formation of a double-bond between two three-fold coordinated silicon
atoms on the surface\cite{honeycomb}. 
In the case of the (3$\times$1) reconstruction a further
stabilization mechanism comes from the fact that,
after the transfer of the valence electrons
from the metal atoms, the  HC structure becomes electronically
closed-shell. \footnote{In the case of the gold induced reconstructions this
last mechanism might be absent or modified
since gold has
a larger electron affinity than silicon and, in principle, the transfer of
electrons should go from silicon to the metal atoms. }
It seems somewhat surprising that the silicon HC structure, 
common to the (3$\times$1) and Si(557)-Au metal induced reconstructions,
is absent from the MP$^+$ model of the
Si(111)-Au-(5$\times$2) surface. 
Indeed the MP$^+$
model seems to be based on an
almost unreconstructed Si(111) 
surface with a row of adatoms on top, and the more 
clear
disturbance from this bonding pattern being the presence of a
surface dislocation.

We now proceed further with the structural relaxations of the MP$^+$ 
system. 
It is instructive to focus first on a optimization were
only the silicon degrees of freedom are taken into account. The gold atoms are
constrained to remain at their initial coordinates. Due to the more directional
bonding of silicon we can expect the structural changes to be simpler to analyze
and somewhat less dependent on the particular choice of the initial guess in
this case. Besides, as a stronger scatterer, we can assume that the gold
positions to be better resolved in the experiment. The resulting geometry is
plotted in Fig.~\ref{fig:mp_plus}(a).
We observe two main effects. On the one hand, the HC
configuration clearly emerges. One of the driving forces behind this 
rebonding is the movement forward of $a$ and $a^\prime$ atoms in order to 
form an additional covalent bond with the silicon atoms in the underlying layer.
The double-bonded ``dimers" of the HC structure are 
formed by atoms $b$ and $c$.
This questions 
the location of the adatoms in the surface since, in principle, 
the dangling-bond associated with atom $b^\prime$
could disappear with the formation of a silicon double bond.
We can observe the disturbing
effect
of the adatom on the HC 
structure.
The appearance of the HC bonding pattern during
the relaxation of the MP$^+$ structure
confirms the results
of recent density functional calculations 
by Kang and Lee~\cite{5x2_comp}, who also made a geometrical
optimization
of the MP$^+$ model.
The electronic bands calculated for this structure 
(not shown here) are also in quite good agreement
with those presented by these authors in 
Ref.~\onlinecite{5x2_comp}.
Fig.\ref{fig:mp_plus} (a) also shows clearly  
what could be classified as a ``stacking fault" in the structure
(bonds 
of atoms $d$ and $e$
coincide with those in the underlying silicon layer).   
This stacking fault, which probably is energetically   
unfavorable, can be easily avoided by moving the position of
the surface dislocation from the right-hand 
to the left-hand of the gold trench. Alternatively we can visualize 
this change (at least approximately)
as a 180$^o$ rotation of the surface layer with respect
to the underlying silicon structure.
This transformation gives one of the structures 
discussed in the next section, which incidentally is
almost identical to the ``5$\times$1" structure
proposed recently by Erwin in Ref.~\onlinecite{erwin03}.

When the relaxation of the MP$^+$ 
system is continued without any constraints,
the monatomic gold wires are strongly distorted
as can be seen in Fig.~\ref{fig:mp_plus} (b). 
This distortion was not
observed in the density functional
calculations of Kang and Lee\cite{5x2_comp}. 
The reasons for this discrepancy 
are not completely clear at the moment.
The break of the monatomic gold wires seems to be related with the
presence of adatoms. If they are eliminated from the structure
the gold atoms remain in two well separated parallel wires.
Additionally, the strain introduced by the adatoms in the structure, 
results in the 
weakening of some Si-Si bonds in the surface layer 
(see the increased distance between atoms $f$ and $g$). 
In spite of these strong structural distortions, the presence 
of adatoms in the structure is still energetically favorable as can
be seen in Table~\ref{tab:systems}.
These structural distortions are reflected in the 
band structure:
following the nomenclature used in
Ref.~\onlinecite{5x2_comp}, the band S$_1$ is 
shifted to higher energies respect to the S$_2$ and 
a gap of $\sim$0.3~eV is opened respect
to the constrained case.

In summary, our results suggest that neither 
the silicon structure nor the positions of the 
gold atom 
in the structure proposed by Marks and Plass~\cite{marks_plass} 
are stable. Furthermore, in agreement
with the general conclusions of Ref.~\onlinecite{5x2_comp}, neither the 
STM images nor the band structure of the fully relaxed or the 
constrained relaxed MP$^+$ model seem to be in
agreement with the experimental information. 
 
\subsection{The Erwin models}

As discussed in the previous section, 
the MP model of the Si(111)-(5$\times$2)-Au 
surface reconstruction is characterized
by the presence of a surface dislocation between
one of gold wires and the neighboring silicon atoms. 
Other locations are possible for the surface dislocation.
In particular, it can 
be translated to the {\it other} gold wire, this 
can also be assimilated
to a rotation of the surface bilayer 
with respect to the underlying
bulk silicon. This eliminates
the ``stacking fault" commented in the previous section, 
and produces a new
structural model. This structure is 
very similar to the ``5$\times$1" model recently proposed
by Erwin\cite{erwin03}, and we refer to it as E(5$\times$1). 
In Table~\ref{tab:systems}
we can see that the E(5$\times$1) model without 
silicon adatoms is slightly more
stable than the relaxed MP structure.
Fig.~\ref{fig:E5x1}~(a) shows the 
relaxed structure of the E(5$\times$1) model.
The left (L) gold wire, where the surface
dislocation is located, suffers a considerable dimerization, which 
is much smaller for the right (R) wire. The alternating Au-Au
distances as obtained with 
VASP are, 4.06~\AA\ and 3.59~\AA\ for the L wire, and 3.82~\AA\ and 
3.83~\AA\ for the R wire. 
The geometries obtained with SIESTA are very similar,
specially those obtained with the more complete DZP basis set. 
Hereafter we name ``SiAu complex" the structure 
formed by the two gold wires and the central silicon atom connecting 
them. The silicon structure in between two of such SiAu complexes
is quite flat
and resembles what could be described as a double honeycomb 
chain (DHC) silicon structure~\cite{erwin03}. 
The band structure along the direction
parallel to the gold wires is shown in Fig.~\ref{fig:E5x1}~(b).
It shows several surface bands and has a metallic character.
Those surface
bands mainly associated with the Si-Au complex 
has been highlighted using solid symbols. Most of these bands
are occupied and appear in the gap region. The unoccupied surface
bands appearing in the gap are mainly associated with the silicon DHC.
The most prominent feature is a dispersive 
band associated with 
the weakly dimerized (right) gold wire and the central silicon atom in the 
SiAu complex. This band is, in principle, metallic and close
to half occupied. Although small gaps are opened associated with the
crossings with other bands and slight geometrical distortions, it
can be easily followed in Fig.~\ref{fig:E5x1}~(b) extending 
from $\sim$1.3~eV below to $\sim$2.3~eV above E$_F$.
A similar band, with a similar origin,
also dominates the band structure
of the Si(557)-Au surface~\cite{daniel1,daniel2}.
This band comes mainly 
from the $sp^3$ lobes of the 
central silicon atom in the SiAu complex. There is also a strong
hybridization with the 6$p$ states of the gold atoms in the R wire.
For this reason, they are better assigned to the Si-Au bonds
connecting the central silicon with the R gold wire. 
Its large dispersion 
is due to the large overlap between these Si-Au bonds 
along the wire. 
The metallicity stems from the inability of gold (each gold atom
only provides one valence 
electron) to 
saturate the bonds with 
all its silicon neighbors~\cite{daniel1}. 
The other states in the Si-Au complex give rise to 
relatively flat surface bands associated either with weakly 
overlapping silicon states or with the gold dimers.

In Ref.~\onlinecite{erwin03} it was also
proposed that, under certain conditions, it could be energetically 
favorable to remove  
some of the over coordinated silicon atoms in the neighborhood
of the surface dislocation. 
Our relaxed structure 
for this model (hereafter E(5$\times$2)) is shown
in Fig.~\ref{fig:E5x2}. In this case both gold wires present an appreciable
dimerization with alternating
Au-Au distances of 4.37~\AA\ and 3.35~\AA\ for the 
left gold wire, and 4.16~\AA\ and 3.49~\AA\ for
the right wire. Our SIESTA calculations with the smaller DZ basis set 
predict the E(5$\times$2) model to be more stable, 
by at least 3.4~meV/\AA$^2$,
than 
both the E(5$\times$1)  model and the different 
variants of the MP model (see
Table~\ref{tab:systems}). However, the difference 
between the E(5$\times$1) and 
E(5$\times$2) models is reduced with the use of more complete basis set.
In particular, our plane-wave calculations predict 
both models to be 
degenerate within 0.1~meV/\AA$^2$ 
(the E(5$\times$1) slightly more stable).
This agrees with the results 
of Ref.~\onlinecite{erwin03} where the E(5$\times$1) model is predicted 
to be more stable than the E(5$\times$2) variant 
by less than 1~meV/\AA$^2$,
and only after the addition of silicon adatoms the E(5$\times$2)
structure becomes favorable.  

The band structure of E(5$\times$2) with zero adatom coverage
is plotted in Fig.~\ref{fig:E5x2B}.
The band structure along
the wires is in good agreement with that reported in 
Ref.~\onlinecite{erwin03} for this structure.  
Again, the surface bands close to the Fermi energy come mainly
from the SiAu complex.  
Like in the case of the E(5$\times$1) model, 
the band structure is
metallic. This is in disagreement 
with one of the latest and more detailed
ARPES experiments which suggests
that the Si(111)-Au-(5$\times$2) surface 
is a semiconductor with a band gap of at least 
0.2~eV~\cite{arpes_latest}. However, 
the metallic versus semiconducting
character of this surface is still a matter of controversy.
For example, 
the recent ARPES study by Himpsel and collaborators finds
several metallic bands~\cite{himpselnewARPES04}. 
In fact, this reference and the scanning 
tunneling spectroscopy (STS) data of Ref.~\onlinecite{Yoon04}
indicate that the surface could be composed
of alternate metallic and semiconducting regions along the 
gold wires. 
Our calculated band structure for the E(5$\times$2) model
is very close to being semiconducting. 
Just by shifting the S$_1$ band to higher energies 
by a few tenths of eV we could obtain a semiconducting
surface. This might indicate that the metallic behavior is simply related 
to the limitations inherent to the local density approximation
used here and the very simplified assumption that the
monoelectronic 
eigenvalues can be directly identified with the photoemission peaks. 
In spite of its metallicity,
several characteristics of the photoemission spectra are
recovered by the band structure in Fig.~\ref{fig:E5x2B}.
The most prominent band observed experimentally 
starts at the boundary 
of the 5$\times$2 zone (ZB$_{\times2}$) dispersing downwards
until it reaches a minimum at the boundary of the 5$\times$1 zone 
(ZB$_{\times1}$)~\cite{5x2_bands,1d_states,arpes_latest,himpselnewARPES04}.
This band appears at binding energies between $\sim$0.2~eV and 
$\sim$1.3~eV.  
Following     
Erwin~\cite{erwin03}, we can try to identify this band with our 
S$_2$ band, whose maximum appears close to E$_F$ in the neighborhood
of ZB$_{\times2}$. 
However, it becomes difficult 
to follow the dispersion of this surface
band 
as we move to higher binding energies for two reasons:
{\it i)} the band enters the region of
the projected
bulk bands, becoming a surface resonance and, {\it ii)}
other surface bands coming from
the 
same region of the surface appear in the energy interval between
-0.5 and -1.2~eV.
This last point
is widely
consistent with the experimental data in Ref.~\onlinecite{arpes_latest}, 
where three additional bands are identified for binding energies
larger than 0.5~eV.

Losio and collaborators~\cite{1d_states} reported an 
interesting effect,
a continuous dimensionality 
transition of the main surface band. The character changes  
from strongly one-dimensional at the band maximum (i.e. only dispersing 
in the direction parallel to the gold wires) to two-dimensional 
at its minimum (i.e. with a non-negligible dispersion also in the 
perpendicular direction). The strong one-dimensional character 
of the surface 
states close to E$_F$ has also been 
confirmed in the most recent ARPES 
measurements~\cite{arpes_latest,himpselnewARPES04}.
The dispersions in the direction 
perpendicular 
to the wires can be found in Fig.~\ref{fig:E5x2B}~(b) and (c).
The band widths
are rather small for most surface bands.
An effect similar to the reported 
dimensionality transition can be seen
in the case of the S$_1$ band. 
It is tempting to assign the experimentally observed 
effect to the S$_2$ band
(see the different dispersion of bands S$_2$ in panel (b) 
and S$^\prime$ in panel (c)).
However, as commented above it is not so simple to 
follow the S$_2$ band as it disperses
downwards. In fact, 
we can locate what seems to be an avoided crossing between 
the S$_2$ and 
the S$_3$ bands half way
along the ZB$_{\times2}$-ZB$_{\times1}$ 
path in Fig.~\ref{fig:E5x2B}~(a). 
Therefore, we think that 
the S$^\prime$ band in panel (c) is rather related to the
S$_3$ band than to the S$_2$ band, and the dimensionality change
would be absent from our results.
Also the energy position of the band S$^\prime$ ($\sim$-0.5~eV) 
is quite far 
from the $\sim$-1.3~eV found experimentally for the band minimum.
Therefore, in contrast to Erwin~\cite{erwin03}
we conclude that our calculated
band structure for the E(5$\times$2) model does not
provide 
a direct explanation
to the observation by Losio {\it et al.}.

Similarly to the surface bands of the E(5$\times$1) model, 
the S$_1$, S$_2$ bands in Fig.~\ref{fig:E5x2B}~(a)
have the largest weight in the central Si atom 
in the SiAu complex.
The S$_1$ band can be associated with SiAu bonds connecting the 
central Si with the left gold wire. 
This SiAu bonds have a small overlap and this is translated in 
a quite flat band.
The two dispersive S$_2$ and S$_3$ bands 
have a stronger weights in the other SiAu bonds, 
which have a larger overlap and, 
therefore, present a stronger dispersion.

We now explore the role of the silicon adatoms in these structures.
We first studied the stability of 
the adatoms 
in the E(5$\times$1) model when they are located over the 
silicon part of the surface reconstruction, i.e on sites equivalent
to those occupied by the adatoms in the original MP proposal. 
It is interesting to note that the role of the silicon 
adatoms in such positions is indeed not very clear.
The stability of the adatoms in typical silicon reconstructions
stems from the fact
that each adatom can saturate three dangling bonds
in the surface at the expense of creating just an additional 
dangling bond. The energy gained in this process
usually overcomes the strain energy caused by the addition 
of the adatoms.
However, the E(5$\times$1) model in Fig.~\ref{fig:E5x1}~(a)
does not have silicon dangling bonds.
The appearance of unsaturated dangling bonds is avoided
by the formation of the double-bonded silicon dimers that 
characterize the HC configuration. In fact, the only metallic band
in this model comes from the SiAu complex as explained above.
In accordance with these observations, we found extremely 
difficult
to reach a stable configuration, i.e. with all the 
components of the forces below our threshold, for 
the silicon adatoms over the silicon sites of the E(5$\times$1) model.
Finally, after several hundreds of optimization steps this
model spontaneously relaxed into a new structure.
This structure, labeled
N$^+$ in Table.~\ref{tab:systems}, belongs to a new family 
of structural models for the Si(111)-Au-(5$\times$2) surface
found in this work for the first time and described 
in more detailed in the next subsection.

In the light of the previous comments, a 
more stable
adsorption site for the silicon adatoms would be
on top of the SiAu complex. This has been 
previously proposed by Erwin~\cite{erwin03},
and is confirmed by our calculations. 
Table~\ref{tab:systems} shows the changes 
in surface energy after the addition of one
silicon adatom per 5$\times$2 unit cell.
The behavior is opposite for the E(5$\times$1) and
E(5$\times$2) models, with the addition being 
energetically favorable for the later model.
The E(5$\times$1) remains metallic after 
the addition of the adatom, and the dispersive
band associated with the SiAu complex remains
quite unchanged. The situation with the E(5$\times$2)
model is different. In
agreement with the results in Ref.~\onlinecite{erwin03}
we find that the band structure becomes semiconducting
after the addition of the adatoms. The corresponding
atomic and electronic structure can be found in 
Fig.~\ref{fig:E5x2star}~(a) and (b) respectively.
The surfaces bands with a larger contribution from the atoms
in the SiAu complex has been highlighted using 
solid symbols. It has been impossible to identify 
a band that can be solely assigned to the adatoms.
We can see that the band structure 
of the E(5$\times$2) suffers major modifications after
the addition of adatoms, at least for the 
large concentrations considered here.
Besides the fact 
that the structure becomes semiconducting, the agreement with 
the detailed photoemission experiments of 
references~\onlinecite{arpes_latest} and
\onlinecite{himpselnewARPES04}
seems to be somewhat degraded.

\subsection{New structural model }

In this section we present a novel structural model for the 
Si(111)-(5$\times$2)-Au surface reconstruction that has been 
found during our investigation. Our slab spontaneously 
relaxed to this new structure while trying to optimize
a modified version 
of the E(5$\times$1) model commented in the previous section.
The new structure can be found in Fig.~\ref{fig:NZERO}, 
and will be referred here as model N.
Table~\ref{tab:systems} shows that the energy of the 
new model compares favorably with 
those of the other structures proposed to date. In fact, 
within our calculational scheme it is the most favorable 
structure. The difference with the second most stable model 
without adatoms, the E(5$\times$2),
is  4.7~meV/\AA$^2$. This difference is reduced to 
4.1~meV/\AA$^2$ when using a more complete DZP basis set
as shown in Table~\ref{tab:systems2}. These 
energy differences
are quite small, so further studies have been performed
in order to drive more definitive conclusions. 
First, we have repeated our calculations 
using the PBE~\cite{gga} GGA exchange and correlation 
functional instead of LDA. 
The new structure 
continues to be more stable by
3.6 and 5.1~meV/\AA$^2$ using, respectively, a DZ and a DZP
basis set. 
As a second step, the energy ordering between 
the N and the E(5$\times$2)
structures has been confirmed using VASP and slabs 
containing three and four silicon double-layers. The new
model is more stable than the E(5$\times$2) 
by at least 2.6~meV/\AA$^2$. 
These results convincingly establish, at least within
the framework of density functional calculations,
the larger 
stability of our new structural model compared to previous proposals
in the limit of negligible adatom 
coverage

Given the small energy differences between both models it may be 
interesting to estimate the effect of the 
vibrational degrees of freedom in the surface free energy
$\gamma(T)$. The
vibrational contribution can affect the energy ordering 
even at zero temperature
due to the zero-point energy, and its importance grows with
temperature. Unfortunately, an accurate 
estimation of the vibrational surface free 
energy $\gamma_{vib} (T)$
is a formidable task that would require the detailed calculation 
of the dynamical properties (phonon band structure)
of the different surface models. This is 
a computationally very demanding calculation that is beyond
the scope of the present paper. We can obtain a rough 
estimation of the 
vibrational contribution to the difference of the surface
free energies between the different structures $\Delta \gamma (T)$
following Ref.~\onlinecite{Reuter02}.
We have
$\Delta \gamma (T)=\Delta E_{surf}+\Delta \gamma_{vib} (T)$, 
where $\Delta \gamma_{vib} (T) \approx$
$3N^{E(5\times2)}_{Si}[F(T,\omega^{E(5\times2)}_{surf})- 
F(T,\omega_{bulk}) ] $  - $
3N^{N}_{Si}[F(T,\omega^{N}_{surf})-
F(T,\omega_{bulk})] $ + $
3N_{Au}[ F(T,\omega^{E(5\times2)}_{Au}) 
-F(T,\omega^{N}_{Au})]$. Here $\Delta E_{surf}$ is given in 
Table~\ref{tab:systems} and is independent of the temperature 
$T$; $N^{E(5\times2)}_{Si}$ and $N^{N}_{Si}$ are
the number of silicon atoms per unit cell in both surface
structures, and $N_{Au}$ the number of gold atoms; $F(T,\omega)$,
given in the Appendix of Ref.~\onlinecite{Reuter02},  
is the free energy of a given vibrational mode $\omega$; 
the frequencies $\omega^{N}_{surf}$, $\omega^{E(5\times2)}_{surf}$, 
and $\omega_{bulk}$
characterize the average 
vibrational properties
of the silicon atoms in 
both surface structures and in bulk silicon, while 
$\omega^{E(5\times2)}_{Au}$ and $\omega^{N}_{Au}$ those of the 
gold atoms in both surfaces.
We take 
for $\omega_{bulk}$ values in the range of 50-70~meV, 
and $\omega_{surf}$ ranging from 
0.5 to 1.5 the $\omega_{bulk}$ value. Within these range of 
parameters, if  $\omega^{N}_{surf}$
and $\omega^{E(5\times2)}_{surf}$ differ less than 
a $\sim$10\%, then
$\Delta \gamma_{vib} (T)$ 
stays within $\sim\pm$2~meV/\AA$^2$ for temperatures up to 300~K. 
If the vibrational
properties of both surface models differ more significantly,
then $\Delta \gamma_{vib} (T)$ can affect the 
relative order of the structures at much lower temperatures.
However, we should not expect strong differences in 
the {\it average} vibrational frequencies of 
the E(5$\times$2) and N models. Both models present
very similar bonding patterns and structures.
It is interesting to notice that 
$\Delta \gamma_{vib} (T)$ is nonzero even if the vibrational
properties of both structures are identical, i.e. 
$\omega^{E(5\times2)}_{surf}$=$\omega^{N}_{surf}$
=$\omega_{surf}$
and $\omega^{E(5\times2)}_{Au}$=$\omega^{N}_{Au}$.
This reflects
the different number of silicon atoms
in the unit cell of the two surface reconstructions.
In this case we have 
$\Delta \gamma_{vib} (T)\approx$ 
$3(N^{E(5\times2)}_{Si}-N^{N}_{Si})[F(T,\omega_{surf})-
F(T,\omega_{bulk})]$. Using the same parameters as above we obtain 
$\Delta \gamma_{vib} (T)$
within $\pm$1.5~meV/\AA$^2$ up 
to $\sim$1000~K. 
Thus we can conclude that
the energy ordering obtained in the
present total energy 
calculations is not altered by the vibrational
contribution to the free energy up to, at least, room temperature.

In the new structure the gold wires
along the $[\bar{1}10]$ direction present 
a dimerization comparable to the E(5$\times$2) structure.
The alternating
Au-Au distances are 3.24~\AA\ and 4.40~\AA\ 
(3.19~\AA\ and 4.45~\AA)
along the right (left) wires.  
The distance between nearest neighbor Au wires 
along the $[11\bar{2}]$ direction is smaller 
in the N structure  (3~\AA) than in the
E(5$\times$2) structure (3.8~\AA). The later value
being in better agreement with the 
$\sim$3.9~\AA\ deduced from the 
HREM studies of the surface.~\cite{marks_plass}

Similarly to the E(5$\times$1) structure, 
most of the surface of the N model is covered with 
a silicon 
double honeycomb chain structure~\cite{erwin03}. 
One of the silicon atoms in the DHC appears at
a higher position over the surface. This indicates
that this atom has a charged dangling-bond and, therefore, 
is trying to  develop a $sp^3$ hybridization.  
This atom is expected to be more visible in the 
STM images and to provide a preferential site for adsorption
on the surface, in particular for possible silicon adatoms.
The boundaries between
the DHC stripes are occupied by the SiAu complex,
in which
a central silicon atom appears bonded
with three gold dimers.

The band structure of the new structure
is plotted in Fig.~\ref{fig:NB}.
The general features are in good
agreement with 
the most recent ARPES 
studies~\cite{arpes_latest,himpselnewARPES04}, although 
some of the details are different.  
The most dispersive and prominent
surface bands are quite similar to those found for the 
E(5$\times$2) model.
The surface is predicted to be semiconducting, which agrees
with the results of Ref.~\onlinecite{arpes_latest}. 
The bands named S$_1$ and S$_2$
by Matsuda {\it et al.}~\cite{arpes_latest} 
can be easily identified
in our calculation, and we use the same notation.
Other less dispersive surface bands are also
observed in our calculated band structure. 
These can be tentatively identified with those labeled S$_3$ and S$_4$ by 
Matsuda {\it et al.}. However, the S$_3$ band appears shifted
to lower binding energies by a few
tenths of eV. We can relate this upward energy 
shift to the use of the LDA in our calculations, which 
is likely to be less suited to describe 
more localized (less dispersive) 
states. Besides this energy shift, the sole major discrepancy
with the experimental band structure in 
Ref.~\onlinecite{arpes_latest} is
the absence of the S$^{\prime}_3$ band. However, this band is not so
clearly resolved in the experiments as the others. 

Different symbols are used in Fig.~\ref{fig:NB} according 
to the main atomic character of the bands.
S$_1$ and S$_2$ come from the Si-Au bonds
in the surface (solid symbols). 
This is common to most of the 
models studied in this paper: the most dispersive surface bands
always originate in the Si-Au bonds, 
with the main character corresponding 
to the 3$p$ states of the central Si atom,
and a strong hybridization with 
the 6$p$ states of the neighboring Au atoms.
The flat S$_3$ band corresponds to the silicon dangling bonds in the middle
of the DHC structure (open triangles). 
The also quite flat S$_4$ band is mainly associated  
with the bonds between the gold atoms and the silicon atoms in the border
of the DHC structure (open squares).
We find several unoccupied surface bands whose atomic character is
difficult to determine.
One of these bands is located at energies very close to E$_F$, 
particularly 
near the $\Gamma$ point. The metallic/semiconducting 
character of the surface is thus governed by the position of this band.
This situation is very similar to that already observed
for the E(5$\times$2) model, although in this case the 
band reaches to lower energies and becomes partially occupied
driving the system to metallic.

In agreement with experiment,
most surface bands show a strong 1D 
character in our new structural model as
can be seen in Fig.~\ref{fig:NB}~(b) and (c).
This is particularly clear in panel (b), where most states are located
within the bulk gap.  In the region displayed in 
panel (c) (at the zone boundary of
the 5$\times$1 Brillouin zone) the S$_2$ and S$_1$ 
bands merge with the 
bulk bands, becoming  surface
resonances. It is no longer possible to identify the S$_1$ and S$_2$ 
resonances
with a single band 
of our finite slab and, as a consequence, it is difficult to follow
the band dispersion of these spectral features
in the direction perpendicular to the gold
wires. However, from the data in panel (c) it is clear that the 
combined effect of the possible dispersion, plus the 
broadening of the resonances
extends over a range of $\sim$0.2~eV, much larger than its
dispersion for energies closer to E$_F$.
This is broadly consistent with the 1D to 2D transition reported
in Ref.~\onlinecite{1d_states}
for the most prominent photoemission feature
as the binding energy increases.

We now explore the structure and energetics of the model N 
under the addition of one
silicon adatom per 5$\times$2 unit cell.
We tried several different adsorption sites:
directly on the SiAu wire following the proposal 
by Erwin~\cite{erwin03} (referred as N$^\star$), 
and bonded to the prominent dangling bond
in the DHC structure occupying 
hollow H$_3$ (N$^+$) or top T$_4$ (N$^{+\prime}$) sites~\cite{Northrup86}. 
As shown is Table~\ref{tab:systems},
this high coverage of adatoms is energetically unfavorable in
all cases by at least 2.7~meV/\AA$^2$. 
This is in contrast with the situation 
for the E(5$\times$2) model, where the addition
of one silicon adatom per unit cell is slightly favorable.
In the N$^\star$ structure (not shown) the silicon adatoms
tend to locate in a peculiar bridge 
position between two gold dimers along the $[\bar{1}10]$ direction.
The structure of the N$^+$ model 
is shown in Fig.~\ref{fig:N}~(a). The silicon atoms bonded to 
the adatom adopt a typical silicon 
configuration although, contrary to what is observed
for the clean Si(111) surface, the hollow site is preferred
over the top site~\cite{Vanderbilt}.

The band structure of the N$^+$ surface is shown
in Fig.~\ref{fig:N}~(b). It is very similar to that found for 
the model without adatoms. 
The S$_1$ and S$_2$ are largely unchanged, which 
clearly indicates its origin in the SiAu complex.
The flat S$_3$ band disappears from the gap region
as a consequence of the saturation 
of the dangling bond with the adatom. A new
unoccupied band, associated with the adatoms, appears instead.
This new band can be found around
$\sim$0.6~eV above
E$_F$ in Fig.~\ref{fig:N}~(b).

\subsection{Adatom coverage}

So far we have only considered the limiting 
cases with zero or maximum 
adatom coverage, which correspond to a number $x$ of 
silicon adatoms per 5$\times$2 unit cell equal, respectively, 
to 0 and 1. 
However, the experimental evidence indicates that the equilibrium
concentration is $x\sim 1/4$, corresponding to 
one adatom per 5$\times$8 supercell.
Under silicon rich conditions the adatom coverage
can be increased in the experiment 
only up to $x\sim1/2$, consistent with a 5$\times$4 periodicity.
We have performed explicit calculations for $x=1/2$, $x=1/3$, and 
$x=1/4$
for our 
two most stable models of the reconstruction in order
to simulate these situations that can be reached experimentally.
Due to the very large supercells necessary 
for these calculations (up to 273 atoms),
we have performed them with the SIESTA 
code and restricted to the use of a 
DZ basis set for silicon.
The results of the energetics as a function
of the adatom content can be found in Table~\ref{tab:systems3} and 
in Fig.~\ref{fig:adatoms}. 
The behavior is opposite for both models, N and E(5$\times$2).
It should be kept in mind that model N favors the adatoms
in hollow sites over the silicon surface, while in 
the E(5$\times$2) structure
the adatoms sit on the gold chains are more favorable.

The surface energy
monotonously decreases as a function of the number of adatoms
for the E(5$\times$2) model.
We do not find any evidence of an energy minimum as a function of
the adatom concentration.
This is in contrast with the suggestion made 
by Erwin in Ref.~\onlinecite{erwin03}. In that reference
the addition of adatoms was studied
using the following simplification:
it was assumed that the sole effect of the adatoms
is to dope the gold chains 
with electrons and the energy of the system
was studied as a function of the doping. Erwin found a 
minimum of the total energy 
for 0.5 extra electrons per 5$\times$2 unit cell.
Since each adatom was found to donate two electrons 
to the surface, this would correspond to the observed
adatom concentration at equilibrium of $x\sim1/4$.
However, our simulations introducing explicitly the 
adatoms in the structure do not confirm this behavior.
The surface energy of the E(5$\times$2) structure
always decreases as the adatom concentration is increased. 
However, the slope
of the curve becomes very small 
for intermediate adatom concentrations, showing a weak
dependence of the surface energy in that region. 
We cannot completely rule out the presence of a minimum
for the surface energy at very low adatom concentrations. However, 
it seems quite unlikely looking at Fig~\ref{fig:adatoms}.
It could also be argued that the DZ basis set is not flexible
enough to produce the correct behavior. 
This seems quite improbable 
looking at the data in Table~\ref{tab:systems2}, which 
clearly show that the energy changes induced by the addition
of adatoms are weakly dependent on the details of the 
calculation.

In the case of the new model N, 
the Fig.~\ref{fig:adatoms} shows that the surface energy systematically
increases as a function of the adatom concentration.
With the DZ basis set the N model is always more stable
than the E(5$\times$2) structure. Using a more complete basis 
set and a converged plane-wave calculation 
we find a crossing: the new model is
always more stable at low adatom coverage, but becomes
unstable compare with the Erwin ``5$\times$2" model 
at larger coverages.
Scaling the data calculated with the DZ basis set to reproduce
the VASP results at the end points (i.e. $x=0$ and $x=1$)
we can estimate that the crossing occurs at $x\sim$1/2.
We can conclude then that the N model is, at least
in the framework of density functional calculations,  
more favorable than the E(5$\times$2) model
for adatom concentrations below $\sim$1/2 adatoms 
per 5$\times$2 cell.

\subsection{Simulated STM images}

The STM images of the Si(111)-(5$\times$2)-Au surface
are characterized 
by the presence of bright ``protrusions" 
and ``Y"-shaped features with a definite orientation 
respect to the underlying lattice~\cite{stm_5x2_2,Yshaped,Yoon04}.
It seems quite well established that the protrusion correspond 
to silicon adatoms~\cite{memory,adatoms,5x2_corr}. However, 
the origin of the 
``Y"-shaped
features is less clear.

Figures~\ref{fig:E5x4STM} and \ref{fig:N5x4STM} 
present our simulations of the STM images  
for the E(5$\times$2)
model at -0.8~eV sample bias and 
the N model at -0.6~eV, respectively.
The simulations have been performed for a 5$\times$4
arrangement of the silicon adatoms, corresponding to 
concentration of adatoms that can be actually reached 
in the experiment. 
In agreement with Ref.~\onlinecite{erwin03} and the experiments
the silicon adatoms show as very pronounced bright protrusions.
With the adatoms directly sited 
on the gold chains, the bright spots appear in the 
middle of the underlying row structures for the E(5$\times$2)
model. For the N model they appear in a more lateral position.
This seems to be in somewhat better agreement with the experimental
images (see, for example, the Figure~1~(b) 
in Ref.~\onlinecite{5x2_corr}).
``Y"-shaped features can be identified in the simulated
STM images of both 
N and E(5$\times$2) models. The possible candidates 
have been highlighted in the 
Figures~\ref{fig:E5x4STM} and 
\ref{fig:N5x4STM} (see also Ref.~\onlinecite{erwin03}). 
The identification is, however, more clear in the case 
of the less symmetric E(5$\times$2) structure. 

\section{Conclusions}

We have performed a systematic study of different models of 
the Si(111)-(5$\times$2)-Au surface reconstruction by means
of first-principles density-functional calculations using the
SIESTA~\cite{siesta1,siesta2} and the 
VASP~\cite{vasp1,vasp2} codes. 
We start our investigation with the structural
model proposed by Marks and 
Plass~\cite{marks_plass} (MP). 
This is the most detailed model of this surface
reconstruction solely based on experimental 
information to date. 
Therefore, it provides a logical 
starting point for our study.
We have also considered different variants of the relaxed
MP model, including the structures recently proposed
by Erwin~\cite{erwin03}, and a new structure found 
during our simulations. Within the computational schemes
used here this new structure is the
most favorable energetically, at least in the regime of
low concentration
of silicon adatoms.
In general, we find a reasonable agreement between our results
and those of the two existing
theoretical studies of the surface~\cite{5x2_comp,erwin03}.
The energy differences between different models
are quite small, with most structures lying 
in a narrow range of surface energies of less than 10~meV/\AA$^2$
(the estimated error bar for our energies
is of the order of  1-2~meV/\AA$^2$).
This, together with the uncertainties arising from the use 
of the local approximation to the density functional theory,  
make difficult to draw definitive conclusions 
solely
based on the energetics.
The comparison of the calculated band structures and
local density of states, respectively, with the available
ARPES data~\cite{1d_states,arpes_latest,himpselnewARPES04} 
and the 
STM images~\cite{stm_5x2,stm_5x2_2,Yoon04} 
becomes then instrumental in order
to identify the most plausible candidates for the equilibrium 
structure. In the following we summarized 
some our main conclusions:

{\it i)} Like in the case of the
reconstructions formed by the deposition of
gold on stepped 
silicon surfaces~\cite{himpsel_review,daniel1,robinson,daniel2},
the silicon honeycomb chain (HC)~\cite{honeycomb}
structure emerges as a fundamental
building block of the reconstruction.
In agreement with the result of Hang and Lee~\cite{5x2_comp},
the silicon HC is formed spontaneously during the relaxation
of the MP model. The HC is also present in the optimized
geometries of all the other 
structural models considered in our work.

{\it ii)} For the MP model 
we agree with the main conclusions of 
Ref.~\onlinecite{5x2_comp} that
neither the simulated
STM images nor the calculated band structure compare
satisfactorily with the experimental data.

{\it iii)}
We have studied in detail the models proposed  
in Ref.~\onlinecite{erwin03} by Erwin, the E(5$\times$1)
and E(5$\times$2) structures. 
The E(5$\times$1) model is quite
similar to the MP structure: they correspond 
to two possible positions, at opposite sides
of the SiAu complex, of the surface dislocation
present in these structures. The E(5$\times$1) model and 
its E(5$\times$2) variant are energetically degenerate
at zero adatom coverage.
However, these two structures
show a 
different behavior against the addition of silicon adatoms:
it is always unfavorable for the E(5$\times$1) model, while tends to 
increase the stability
of the E(5$\times$2) model. 

{\it iv)}  
We have explored a different position of 
the surface dislocation:
at the center of the SiAu complex. 
We arrive in this way to a new structure, the N model.
According to our calculations this new structure is  
more stable, at least for low coverages of silicon adatoms, 
than any of the models proposed to date.
The distance between the gold wires in this model is
$\sim$3~\AA, which seems somewhat small compared to the 
$\sim$3.9~\AA\
deduced from the HREM measurements~\cite{marks_plass}.

{\it v)} 
The calculated band structures 
of the E(5$\times$2) and N models without adatoms are
quite similar and appear to be 
in reasonable agreement with the available ARPES
data~\cite{1d_states,arpes_latest,himpselnewARPES04}. 
The other models 
fail to reproduce the main features observed experimentally.
The agreement seems to be
particularly good in the case of the N model.
According to our analysis the most prominent and dispersive
surface bands, named S$_1$ and S$_2$ in 
Ref.~\onlinecite{arpes_latest},
come from the atoms in 
the SiAu complex. In the case of the N model 
the silicon adatoms 
tend to adsorb on the silicon part of the surface, i.e. bonded
to three silicon atoms in the surface layer.
As a consequence, 
the topology and the energy position of these bands are quite
insensitive to the coverage of silicon adatoms.
This contrast with the situation found for the E(5$\times$2) model.
Here
the silicon adatoms tend to adsorb directly 
on the SiAu complex, thus causing a notable modification of the 
surface bands that worsens the agreement with the experimental 
ARPES spectra.

{\it vi)} We have studied the energetics of the E(5$\times$2)
and the N models as a function of the concentration of silicon 
adatoms. Contrary to the suggestion 
of Ref.~\onlinecite{erwin03}, we do not find any evidence
of a minimum of the 
surface energy of the E(5$\times$2) model 
as a function of the adatom coverage.
The surface energy  
always decreases with the addition of adatoms,
although the changes
are very small in the range of $x$ between 1/2 and 1/4,
where $x$ is the number of adatoms per 5$\times$2 unit cell.
For the N model the addition of the adatoms is always
unfavorable. As a consequence of this opposite
behavior, the E(5$\times$2) structure becomes more 
stable than the N structure in the limit of relatively
large adatom
concentrations, for x$\gtrsim$1/2 (notice that
$x=1/2$ corresponds to a 5$\times$4 periodicity). According to 
this picture the exact content of adatoms is instrumental
to determine the equilibrium structure of the reconstruction
within the range of experimentally realizable adatom coverages.
This introduces a new degree of complexity 
that should be taken into account
when analyzing the experimental information.
In particular, this might be behind the observed 
phase separation into 5$\times$4 and 5$\times$2 
patches~\cite{5x2_corr,himpselnewARPES04}. 

{\it vii)} The simulated STM images of the most stable
models, N and E(5$\times$2), are in broad agreement
with the experimental images. The silicon atoms 
produce bright spots which are located in the middle
of the underlying row structures for the E(5$\times$2)
and in a somewhat more lateral position for the 
N model. In both cases ``Y"-shaped features similar
to those observed in the experiment can be found. However,
they are more clear in the case of the E(5$\times$2) model~\cite{erwin03}
where the structure surrounding the gold chains 
is less symmetric.

\begin{acknowledgments}
We want to thank P. M. Echenique 
for continuous support and encouragement. 
This work was supported by the Basque Departamento de Educaci\'on,
the UPV/EHU (Grant No. 9/UPV 00206.215-13639/2001),
the Spanish Ministerio de Educac\'on y Ciencia 
(Grant No. FIS2004-06490-C3-02), and the 
European Network of Excellence FP6-NoE ``NANOQUATA" (500198-2).
S. R. also acknowledges support from the Emil Aaltonen Foundation.
D. S.-P. also acknowledges support from the Spanish Ministerio de 
Educaci\'on y Ciencia 
and CSIC through the "Ram\'on y Cajal" program.

\end{acknowledgments}

\newpage

\begin{table*}[ht]
\begin{tabular}{|l|l|c|}
\hline
Model    & Description     & $\Delta$E$_{surf}$ (meV/\AA$^2$)     \\
\hline\hline
MP$^+$       & Marks and Plass model after a constrained relaxation  & +46.8 \\
RMP$^+$      & Fully relaxed MP$^+$ structure                           & +5.4\\
RMP       & Relaxed MP$^+$ structure without adatoms                 & +8.3 \\
\hline\hline
E(5$\times$1)    & Erwin ``5$\times$1''                                         & +4.8 \\
E(5$\times$1)$^\star$   & E(5$\times$1) with adatoms on the Au-wires              & +6.5 \\
E(5$\times$2)    & Erwin ``5$\times$2'' & +1.4 \\
E(5$\times$2)$^\star$   & E(5$\times$2) with adatoms on the Au-wires              &   0.0 \\
\hline\hline
N         & New model      & -3.3 \\
N$^+$        & N with silicon adatoms in H$_3$ positions        & -0.6 \\
N$^{+\prime}$ & N with silicon adatoms in T$_4$ positions       & +1.2  \\
N$^\star$     & N with adatoms on the Au-wires                   & +2.4\\
\hline
\end{tabular}
\caption{\label{tab:systems}
Summary of the structural models studied here for 
the Si(111)-(5$\times$2)-Au 
reconstruction and their relative 
surface energies ($\Delta$E$_{surf}$).
Those structures containing adatoms have one silicon 
adatom per 5$\times$2 cell, i.e.
the concentration of adatoms is maximum.
Superscript $+$ indicates the presence of ``conventional"
adatoms saturating silicon dangling bonds in the surface.
Labels H$_3$ and T$_4$  refer, respectively, 
to adatoms occupying hollow and top sites~\cite{Northrup86}.
The presence of adatoms 
located on top of the Au wires is 
indicated by a $\star$ superscript. 
The data in this table have been calculated using the SIESTA code with 
a DZ basis for silicon and DZPs-SZd basis for gold. The slabs contained
two silicon bilayers below the surface layer 
(see Fig.~\ref{fig:slab}~(a)). 
All energies
are referred to that of the structure recently 
proposed by Erwin in Ref.~\onlinecite{erwin03}.}
\end{table*}
                                                                                                                            
\begin{table}[ht]
\begin{tabular}{|l|c|c|c|c|}
\hline
Model    & \multicolumn{4}{c|}{$\Delta$E$_{surf}$~(meV/\AA$^2$)} \\
\hline
         & \multicolumn{3}{c|}{SIESTA}& VASP \\
\hline
         &   DZ-3~blys & DZ-2~blys & DZP-2~blys & 2~blys \\ 
\hline\hline
E(5$\times$2)        & +1.3   &  +1.4   &  +1.6 & +1.4  \\
E(5$\times$2)$^\star$ & 0.0   &  0.0   &  0.0  & 0.0 \\
N             & -3.4  & -3.3      & -2.5  & -1.3 \\
N$^+$            & -1.0  & -0.6      &  +0.9 & +1.5  \\  
\hline
\end{tabular}
\caption{\label{tab:systems2}
Convergence of the relative surface energies ($\Delta$E$_{surf}$)
of the most stable structural models 
respect to the basis set and the thickness of the slabs used
in the calculations. The first column shows 
the data obtained with the SIESTA code using a
DZ basis for silicon and three silicon bilayers 
below the surface 
to construct the slab. 
In the second column a slab with only two underlying 
silicon bilayers was used.
The third and four columns are obtained using the thinnest 
slab and, respectively, a DZP basis
set for silicon and the VASP plane-wave code.
}
\end{table}

\begin{table}[ht]
\begin{tabular}{|l|c |c| c| c| c|}
\hline
Model    & \multicolumn{5}{c|}{$\Delta$E$_{surf}$~(meV/\AA$^2$)} \\
\hline\hline
         &          $x=0$      & $x=\frac{1}{4}$ &  
$x=\frac{1}{3}$  
& $x=\frac{1}{2}$ &  $x=1$  \\
E(5$\times$2)$_{x}^\star$   & +1.39 & +0.90 &  +0.86 &+0.81 & 0.0  \\
N$_x^+$        & -3.35 & -1.98 & -1.72  & -1.23  & -0.60 \\
\hline
\end{tabular}
\caption{\label{tab:systems3}
Relative surface energies ($\Delta$E$_{surf}$)
of the most stable structural models as a function of $x$, 
the number
of silicon adatoms per 5$\times$2 unit cell. 
The calculational parameters here are the same used
in Table~\ref{tab:systems}. Notice that 
E(5$\times$2)$_{x=0}^\star$=E(5$\times$2) and
N$_{x=0}^+$=N.
}
\end{table}
\begin{figure*}[ht]
\caption{\label{fig:slab}
(color online). 
Schematic view of a typical slab used in our calculations.
It shows the model proposed
by Marks and Plass (MP$^+$)~\cite{marks_plass} for the
Si(111)-(5$\times$2)-Au surface reconstruction. Large 
circles in the surface layer represent the gold atoms.
The bottom surface of the slab is saturated with hydrogen atoms. (a)
Side view and (b) top view with some of the silicon atoms in the surface
and the two gold chain
labeled (see the text).
}
\end{figure*}

\begin{figure}
\caption{\label{fig:bzs}
(a) Comparison of the bidimensional Brillouin zones corresponding to 
(5$\times$1), (5$\times$2) and (5$\times$4) supercells on the 
Si(111) surface. (b) Schematic view of the path (dotted lines)
used to plot
the band structures in this work ($\Gamma$-
ZB$_{\times2}$-ZB$_{\times1}$-ZB$_{\times2}^{\prime}$-M-$\Gamma$).
Its relation with the (5$\times$1) (dashed lines) 
and (5$\times$2) 
(solid lines) Brillouin zones is indicated, and some special 
points are defined.}
\end{figure}

\begin{figure}
\caption{\label{fig:mp_plus}
(color online). (a) MP$^+$ after the optimization
of the position of the silicon atoms in the structure.
The gold atoms are kept in the positions obtained after the initial
contrained relaxation of the experimental coordinates.
The silicon honeycomb chain (HC) structure has been highlighted. 
(b) The same structure after full relaxation (RMP$^+$).
See the text for the labels of the different atoms.}
\end{figure}

\begin{figure*}
\caption{\label{fig:E5x1} (color online).
(a) Relaxed geometry of the E(5$\times$1) model with 
zero adatom coverage and,  
(b) the corresponding 
band structure. Solid symbols indicate those 
bands with a larger weights in the 
atoms of the SiAu complex. The energies are referred 
to the Fermi level.
}                                                         
\end{figure*}

\begin{figure}
\caption{\label{fig:E5x2} (color online).
Relaxed geometry for the E(5$\times$2) model.}
\end{figure}

\begin{figure*}
\caption{\label{fig:E5x2B}
Band structure of the 
E(5$\times$2) model with zero adatom coverage parallel 
($\Gamma$-ZB$_{\times2}$-ZB$_{\times1}$-ZB$^\prime_{\times2}$ 
path in (a)) and
perpendicular to the gold wires i
($\Gamma$-M path in (a) and panels (b) and (c)). 
Surface bands with the larger contributions coming 
from the atoms in the SiAu complex
are indicated by filled circles.
The energies are referred to the Fermi level.
}
\end{figure*}

\begin{figure*}
\caption{\label{fig:E5x2star} 
(color online).
(a) Relaxed geometry of the E(5$\times$2)$^\star$ model 
(containing one adatom per 5$\times$2 unit cell) and, (b)
the corresponding band structure.
Surface bands with a larger weight in the atoms of 
the SiAu complex are marked with 
filled circles.
The energies are referred to the Fermi level.
}
\end{figure*}
                                                                                
\begin{figure}
\caption{\label{fig:NZERO} (color online).
New structural model for the Si(111)-(5$\times$2)-Au reconstruction.
This is the most stable configuration of the surface according
to our calculations (see Table~\ref{tab:systems}).
}
\end{figure}

\begin{figure*}
\caption{\label{fig:NB}
Band structure corresponding to the N model 
with zero adatom coverage parallel
($\Gamma$-ZB$_{\times2}$-ZB$_{\times1}$-ZB$^\prime_{\times2}$
path in (a)) and
perpendicular to the gold wires ($\Gamma$-M path in (a) and panels (b) and (c)).
Surface bands are marked according to its main atomic character: filled circles indicate
a strong contribution from the atoms in the SiAu complex, open triangles
from the silicon dangling bonds in the
middle of the double honeycomb chain (DHC), 
and open cubes
from those silicon atoms at the boundaries of the DHC stripes, 
neighboring to the gold wires.
The energies are referred to the Fermi level.
}
\end{figure*}

\begin{figure*}
\caption{\label{fig:N} 
(color online).
(a) Optimized geometry of the N$^+$ model (containing one adatom
per 5$\times$2 unit cell) and, (b) the corresponding band structure.
Surface bands with strong contributions from 
the SiAu complex are indicated by filled circles, 
while the states associated to the adatom are marked with open triangles.i
The energies are referred to the Fermi level.}
\end{figure*}

\begin{figure}
\caption{\label{fig:adatoms}
Relative surface energies as a function of the adatom content.
Explicit calculations have been performed for several adatom
concentrations using the smaller DZ basis set (circles). 
The results obtained
with the DZP basis set (diamonds) and 
with plane-wave VASP calculations (triangles) for
the two limiting cases are also shown for comparison. All energies
are referred to those of the E(5$\times$2)$^\star$ model.}
\end{figure}

\begin{figure}
\caption{\label{fig:E5x4STM}
(color online).
Simulated STM image of the E(5$\times$2) model with a sample bias of -0.8~eV
and an adatom concentration corresponding to a 5$\times$4 periodicity.
A possible candidate for the Y-shaped structure is schematically indicated.
The atomic structure is superimposed with the simulated image in the lower
part of the figure. Large circles indicate the positions of Au atoms.}
\end{figure}

\begin{figure*}
\caption{\label{fig:N5x4STM}
(color online).
Same as Fig.~\ref{fig:E5x4STM} but for the new N structure and a sample
bias of -0.6~eV. }
\end{figure*}

\end{document}